\documentclass[journal, twocolumn]{IEEEtran}\IEEEoverridecommandlockouts

\usepackage{amssymb}
\usepackage{amsmath}
\usepackage{amsfonts}
\usepackage{graphicx}
\usepackage{algorithm}
\usepackage{algorithmic}
\usepackage{epstopdf}
\usepackage{cite}
\usepackage{stfloats}
\usepackage{mathrsfs}
\usepackage{multicol}
\usepackage{subfigure}
\usepackage{color}
\usepackage{caption}
\usepackage{setspace}

\newtheorem{theorem}{\textbf{Theorem}}

\newtheorem{definition}{Definition}

\newtheorem{lemma}{\textbf{Lemma}}

\begin{document}
\title{\huge Analysis and Optimization of Successful Symbol Transmission Rate for Grant-free Massive Access with Massive MIMO}
%\author{Gang~Chen,~Ying~Cui,~Hei~Victor~Cheng,~Feng~Yang,~and~Lianghui~Ding
%	\thanks{The authors are with Shanghai Jiao Tong University, China.
%	}
%}
\author{Gang~Chen,~Ying~Cui,~Hei~Victor~Cheng,~Feng~Yang,~and~Lianghui~Ding 
	\thanks{G.~Chen,~Y.~Cui,~F.~Yang,~and~L.~Ding  are with the Department of Electronic Engineering, Shanghai Jiao Tong University, Shanghai 200240, China (e-mail: yangfeng@sjtu.edu.cn).}
	\thanks{H.V.~Cheng is with  The Edward S. Rogers Sr. Department of Electrical
			and Computer Engineering, University of Toronto, Toronto, ON M5S 3G4,
			Canada.}\vspace{-0.7cm}
}
\maketitle
\begin{abstract}
Grant-free massive access is an important technique for supporting massive machine-type communications (mMTC) for Internet-of-Things (IoT). Two important features in grant-free massive access are low-complexity devices and short-packet data transmission, making the traditional performance metric, achievable rate, unsuitable in this case. In this letter, we investigate  grant-free massive access in a massive  multiple-input multiple-output (MIMO) system. We consider random access control, and adopt approximate message passing (AMP) for user activity detection and channel estimation in the pilot transmission phase and small phase-shift-keying (PSK) modulation in the data transmission phase. We propose a more reasonable performance metric, namely successful symbol transmission rate (SSTR), for grant-free massive access. We obtain closed-form approximate expressions for the asymptotic SSTR in the cases of maximal ratio combining (MRC) and zero forcing (ZF) beamforming at the base station (BS), respectively. We also maximize the asymptotic SSTR with respect to the access parameter and pilot length.
\end{abstract}

\section{Introduction}
Grant-free massive access is an important technique for supporting massive machine-type communications (mMTC) for Internet-of-Things (IoT), which is one of the three main use cases for 5G. In grant-free massive access, there are two phases, i.e., the pilot transmission phase and the data transmission phase.  A main technical challenge in massive access is to detect active users and estimate their channels in the pilot transmission phase in the presence of an excessive number of potential users. As only a small subset of users is active at any given time, the user activity detection and channel estimation problem can be modeled as a compressed sensing problem.
Among the existing algorithms for compressed sensing, approximate message passing (AMP) algorithm is widely adopted, as it provides a good tradeoff between performance and complexity. %Authors in \cite{chen2018sparse} first applied AMP in both single antenna and multiple antennas scenarios, it was showen that the MMSE denoiser provides a better performance than other denoisers.
In \cite{liu2018partI,liu2018massive2}, the authors adopt AMP for user activity detection and channel estimation in massive multiple-input multiple-output (MIMO) systems. The asymptotic performance of user activity detection and channel estimation is analyzed in \cite{liu2018partI}, and the asymptotic achievable rate is analyzed in \cite{liu2018massive2} (assuming perfect user activity detection). In \cite{Control}, the authors propose channel-based access control and modified AMP for user activity detection, and analyze the performance of user activity detection.
Note that in \cite{liu2018partI} and \cite{Control}, performance analysis of the data transmission phase is not considered.

Two main features of data transmission in mMTC distinct it from data transmission in traditional human-type communications. Firstly, most data packets are \emph{short}, i.e., usually contain a few bytes. Secondly, low-complexity devices are used, and thus \emph{small modulation} and  \emph{simple channel coding} are preferable. Thus, the achievable rate adopted in \cite{liu2018massive2}, which is an information-theoretic limit in the infinite blocklength regime, may not be a suitable performance metric for data transmission in mMTC. To the best of our knowledge, existing analytical results for data transmission cannot reflect the aforementioned features of mMTC.  In addition, the authors in \cite{liu2018massive2} optimize the pilot length to maximize the achievable rate for only one user activity realization, without considering the activity statistics, making the obtained pilot length less suitable for the case where the total number of active users has a large variance. Finally, the authors in \cite{Control} optimize the access control parameter to maximize the user identification performance,  without considering the channel estimation and data transmission, making the obtained access control applicable only for limited scenarios.

In this letter, we would like to address the above issues. We study grant-free massive access in a massive MIMO system. We consider random access control, and adopt AMP for user activity detection and channel estimation. Considering low-complexity devices, we adopt small phase-shift-keying (PSK) modulation, e.g., BPSK and QPSK, for data transmission according to the standards \cite{NBIoT}. In addition, considering transmission of short data packets, we propose a new performance metric, namely successful symbol transmission rate (SSTR), which reflects the performance of user activity detection and channel estimation in the pilot transmission phase and the performance of detection in the data transmission phase. The proposed SSTR is a more suitable performance metric for mMTC than the achievable rate \cite{liu2018massive2}, and its analysis is also more challenging. We first obtain closed-form approximate expressions for the asymptotic SSTR in the cases of maximal ratio combining (MRC) and zero forcing (ZF) beamforming at the base station (BS), respectively. The analytical results significantly facilitate the evaluation and optimization of the SSTR. Then, we maximize the asymptotic SSTR by optimizing the access parameter and pilot length. The optimization results provide practical guidelines for the design of mMTC systems. Finally, numerical results demonstrate the accuracy of the analysis and the importance of the optimization.
\vspace{-3pt}
\section{System Model} %$n\in\mathcal{N}$
\vspace{-2pt}
Consider a massive access scenario arising from mMTC in a single cell with $N$ users (devices) \cite{liu2018partI,liu2018massive2,EL2018grant}. Let $\mathcal{N}$ denote the set of all users. The BS is equipped with $M$ antennas while each user is equipped with one antenna. We adopt a block-fading channel model where the channels within one coherence interval (CI) of length $T$ symbols remain constant. We consider transmission in one CI, and denote the complex uplink channel vector from user $n$ to the BS by $\mathbf{h}_n\in\mathbb{C}^{M \times 1}$. Assume $\mathbf{h}_n\sim \mathcal{CN}(\mathbf{0},\gamma_n\mathbf{I}_M)$, where $\gamma_n$ represents the path loss and shadowing component \cite{liu2018partI}. Assume that $\gamma_n, n\in\mathcal{N}$ are perfectly known at the BS, and that all users are perfectly synchronized.  We consider random access control with access parameter $\epsilon$. Within each CI, the users generate data with probability $p_a$, and access the channel with probability $\epsilon$ once they have data to send, both in i.i.d. manners. Thus, within each CI, the users send data via the channel (i.e., become active) with probability $p_a\epsilon$ in an i.i.d. manner. Note that  $p_a$ is a given system parameter, and $\epsilon$ is a design parameter for access control (controlling transmitting user sparsity) which will be optimized later. Denote by $\alpha_n\in\{1,0\}$ the random activity state of user $n$ with Pr$[\alpha_n\!\!=\!1]\!=\!p_a\epsilon$.

We adopt a grant-free multiple-access scheme, where each user $n\in \mathcal{N}$ is assigned a unique pilot sequence with $L$ symbols, denoted by $
\mathbf{a}_n \triangleq
\setlength{\arraycolsep}{0.5pt}
( \begin{array}{lll}
a_{n,1}, & \cdots, & a_{n,L}
\end{array} ) \in \mathbb{C}^{L \times 1}
$. The pilot sequences and their correspondence to the user identities are known at the BS.  In a massive access scenario, the pilot length is typically much smaller than the total number of users, i.e., $L\ll N$. Thus, it is not possible to assign mutually orthogonal pilot sequences to all $N$ uses. Note that $L$ is a design parameter which will be optimized later.
As in \cite{liu2018partI,liu2018massive2,Control,EL2018grant}, assume that for all $n\in \mathcal{N}$, the entries of $\mathbf{a}_n$ are independently generated according to $\mathcal{CN}(0,1/L)$. Each CI has two phases which will be illustrated below.
\vspace{-5pt}
\subsection{Pilot Transmission Phase}
\vspace{-3pt}
In the first phase, i.e., the pilot transmission phase, the active users synchronously send their pilot sequences to the BS. Therefore, the matrix of received signals at $M$ antennas  $\mathbf{Y}^{\mathrm{pilot}}\in\mathbb{C}^{L \times M}$ is given by:
\begin{equation}
\setlength{\abovedisplayshortskip}{3pt}
\setlength{\belowdisplayshortskip}{3pt}
\mathbf{Y}^{\mathrm{pilot}}=\sum_{n\in\mathcal{N}}\sqrt{L\rho^{\mathrm{pilot}}_n}\alpha_n\mathbf{a}_n\mathbf{h}_n^T+\mathbf{Z},
\end{equation}
where $L\rho_{n}^{\mathrm{pilot}}$ represents the transmit energy for the pilot sequence of user $n$, and 
$
\mathbf{Z}\in\mathbb{C}^{L \times M}
$
%with $\mathbf{z}_m \sim \mathcal{CN}(\mathbf{0}, \sigma^2 \mathbf{I}),\forall m$
 is the additive noise at the BS with each element following $\mathcal{CN}(0, \sigma^2)$. Denote $\mathbf{x}_n\triangleq\alpha_n\mathbf{h}_n\in\mathbb{C}^{M \times 1},n\in\mathcal{N}$.
The goal of the BS in the pilot transmission phase is to detect user activities and estimate the channels of active users by recovering $\mathbf{x}_n,n\in\mathcal{N}$ from the noisy observations $\mathbf{Y}^{\mathrm{pilot}}$.
As $p_a\epsilon\ll 1$, a lot of $\mathbf{x}_n,n\in\mathcal{N}$ are zero vectors. Thus, such a reconstruction problem is a compressed sensing problem. 
Following \cite{liu2018partI}, this paper adopts a low-complexity AMP algorithm to recover $\mathbf{x}_n,n\in\mathcal{N}$, as it provides a good tradeoff between performance and computational complexity. For all $n\in\mathcal{N}$, based on the estimate $\hat{\mathbf{x}}_n$ of $\mathbf{x}_n$, the detected user activity $\hat{\alpha}_n\in\{0,1\}$ can be obtained by hard-decision detection, and if $\hat{\alpha}_n=1$, the estimated channel vector $\hat{\mathbf{h}}_n$ for $\mathbf{h}_n$ is $\hat{\mathbf{x}}_n$. Denote $\Delta{\mathbf{h}}_n$ as the corresponding channel estimation error for each user $n$, i.e., $\mathbf{h}_n=\hat{\mathbf{h}}_n+\Delta{\mathbf{h}}_n$.
Moreover, the convergence results of AMP provide the distributions of the estimates $\hat{\mathbf{x}}_n,n\in \mathcal{N}$ and estimation errors $\Delta{\mathbf{x}}_n\triangleq\mathbf{x}_n-\hat{\mathbf{x}}_n,n\in \mathcal{N}$.
\vspace{-5pt}
\subsection{Data Transmission Phase}
\vspace{-3pt}
In the second phase, i.e., the data transmission phase, the active users directly send their data to the BS using the remaining $T-L$ symbols. We adopt PSK modulation for data transmission, e.g., BPSK and QPSK, as suggested in the standards \cite{NBIoT}. Let $s_n^W$ denote a  $W$-array PSK symbol of user $n$ with unit power, i.e., $\|s_n^W\|^2=1$, where $W\in \{2,4,\cdots\}$. 
%If $\alpha_n=0$, $s_n^W$ is void. 
Therefore, the received signal at the BS is expressed as:
\begin{equation}\label{eq:data}
\setlength{\abovedisplayshortskip}{1pt}
\setlength{\belowdisplayshortskip}{1pt}
	\mathbf{y}^{\mathrm{data}}=\sum_{n\in\mathcal{N}:{\alpha}_{n}=1}\sqrt{\rho_{n}^{\mathrm{data}}}{\mathbf{h}}_{n} s_{n}^W + \mathbf{z}^{\text{data}},
\end{equation}
where $\rho_n^{\mathrm{data}}$ represents the transmit power for a data symbol of user $n$, and $
\mathbf{z}^{\text{data}}\in\mathbb{C}^{M \times 1}
$
is the additive noise at the BS with each element following $\mathcal{CN}(0, \sigma^2)$.

Based on the detected user activities and estimated channels, the BS tries to decode the data symbols of the users that are detected to be active using two linear receive beamforming strategies, namely MRC and ZF. Denote:
\begin{equation}\label{beamforming}
\setlength{\abovedisplayshortskip}{1pt}
\setlength{\belowdisplayshortskip}{1pt}
\hat{\mathbf{U}}^i\triangleq\left\{
\begin{array}{ll}
\!\!\hat{\mathbf{G}},&i=\text{MRC}\\
\!\!\hat{\mathbf{G}}\left(\hat{\mathbf{G}}^H\hat{\mathbf{G}}\right)^{-1},&i=\text{ZF}
\end{array}\right.,
\end{equation}
where $\hat{\mathbf{G}}\!\triangleq\!(\hat{\mathbf{h}}_n)_{n\in\mathcal{N}:\hat{\alpha}_n=1}\!\in\!\mathbb{C}^{ M\times \hat{K}}$ with $\hat{K}\!\triangleq\! \sum_{n\in\mathcal{N}}\hat{\alpha}_n$ denoting the number of the users that are detected to be active. Let $\hat{\mathbf{u}}_n^{i}$ denote the column of $\hat{\mathbf{U}}^i$ that corresponds to user $n$ with $\hat{\alpha}_n=1$.
Employing beamforming vector $\hat{\mathbf{u}}_n^{i}$, by (\ref{eq:data}) and $\mathbf{h}_n=\hat{\mathbf{h}}_n+\Delta{\mathbf{h}}_n$,  we have:
\begin{align}
\hat{r}^{i,W}_n
&=\hat{\mathbf{u}}_n^{iH}\mathbf{y}^{\mathrm{data}}\nonumber\\
&=\hat{\mathbf{u}}_n^{iH}\left(\sum_{n\in\mathcal{N}:{\alpha}_{n}=1}\sqrt{\rho_{n}^{\mathrm{data}}}\left(\hat{\mathbf{h}}_n+\Delta\mathbf{h}_n\right) s_{n}^W + \mathbf{z}^{\text{data}}\right)\nonumber\\
&=\sqrt{\rho_n^{\mathrm{data}}}\hat{\mathbf{u}}_n^{iH} \hat{\mathbf{h}}_n s_n^W  + \hat{\mathbf{u}}_n^{iH}\!\!\!\! \sum_{n'\in\mathcal{N}:{\alpha}_{n'}=1, n'\neq n}\!\!\!\!\sqrt{\!\rho_{n'}^{\mathrm{data}}}\hat{\mathbf{h}}_{n'} s_{n'}^W\nonumber\\ 
&+\hat{\mathbf{u}}_n^{iH} \sum_{n'\in\mathcal{N}:{\alpha}_{n'}=1}\sqrt{\rho_{n'}^{\mathrm{data}}}\Delta\mathbf{h}_{n'} s_{n'}^W  + \hat{\mathbf{u}}_n^{iH} \mathbf{z}^{\text{data}}.\label{beam}
\end{align}
Then, the BS performs the minimum-distance detection on $\hat{r}^{i,W}_n$ by treating the term induced by channel estimation errors and interference from other users as additional noise, and obtains the estimated symbol $\hat{s}^{i,W}_n$ for user $n$ with $\hat{\alpha}_n=1$.

\section{Performance Metric}
In this letter, we use the SSTR, which represents the total number of symbols that can be correctly detected at the BS within a CI, as the performance metric for data transmission in grant-free massive access.

\begin{definition}\label{theorem:1}
	For given pilot length $L$ and access parameter $\epsilon$, the SSTR under the receive beamforming strategy $i$ and the PSK modulation of size $W$ is defined as:
%	\begin{equation}
%	\Phi^{i,W}\triangleq\frac{T-L}{T}\mathbb{E}\left[\sum_{n\in\mathcal{K}}\left(1-\psi^{i,W}_{n}\right)\left(1-P_{n}^{MD}\right)\right],\label{eq:str}
%	\end{equation}
\begin{equation}\label{eq:def}
\setlength{\abovedisplayshortskip}{1pt}
\setlength{\belowdisplayshortskip}{1pt}
	\Phi^{(i,W)}(L,\epsilon)\!\triangleq\!\frac{T\!-\!L}{T}\mathbb{E}\!\left[ \sum_{n\in\mathcal{N}} \!\mathrm{I}[\alpha_n\!=\!1,\hat{\alpha}_n\!=\!1,\hat{s}_n^W\!\!=\!s_n^W] \right]\!,\!\!
\end{equation}
where $\mathrm{I}[\cdot]$ represents the indicator function, and the expectation is taken over all sources of randomness.
\end{definition}

Note that the SSTR captures user activity detection errors, channel estimation errors and data detection errors. %Note that although the channel estimation errors are not present in the formula, it affects the SER and thus the SSTR in an indirect way as we will set forth below.
The SSTR is a more suitable performance metric for grant-free massive access. However, in the general case, the analytical form of $\Phi^{(i,W)}(L,\epsilon)$ is not tractable, due to the complicated signal processing in grant-free massive access. Thus, as in \cite{liu2018massive2}, we focus on the asymptotic case. Specifically, in Section III and Section IV, we consider the asymptotic analysis and optimization of the SSTR at large $M, N$ and $L$ and high SNR under a simple power control policy, i.e., statistical channel inversion, which can reduce the channel gain differences between users, and is especially beneficial to users with relatively weaker channel gains \cite{EL2018grant}. 

With statistical channel inversion, $\rho_n^{\mathrm{pilot}}, n \in \mathcal{N}$ and $\rho_n^{\mathrm{data}}, n \in \mathcal{N}$ are chosen such that 
$\rho_n^{\mathrm{pilot}}\gamma_n\!=\!\rho_n^{\mathrm{data}}\gamma_n\!=\!\gamma, n\! \in\! \mathcal{N}$, 
where $\gamma$ denotes the receive power for both pilot symbols and data symbols at each user. That is, the transmission powers of users scale inversely proportionally to their path-loss and shadowing components. With the same receive power, all users are statistically the same.
Therefore, we can drop the user index $n$, and some dependence on  $(\alpha_n)_{n\in{\mathcal{N}}}$ reduces to the dependence on the number of active users $K\!\!\!\triangleq\!\!\sum_{n\in\mathcal{N}}\alpha_n$. Note that $K$ follows binomial distribution $\mathcal{B}(N,p_a\epsilon)$, i.e., 
\begin{equation}
\setlength{\abovedisplayshortskip}{2pt}
\setlength{\belowdisplayshortskip}{2pt}
\mathrm{Pr}[K\!\!=\!k]\!=\!C_N^k (p_a\epsilon)^k (1\!-\!p_a\epsilon)^{N-k}\!\!\triangleq\! q(N,\!k),
\end{equation} 
where $k\!=\!0 \cdots N$. When there are $k$ active users and the pilot length is $L$, all $k$ active users have the same average probability of missed detection, denoted by $p(k,\!L)\!\triangleq\!\mathbb{E}_{\mathbf{H}}[\mathrm{Pr}[\hat{\alpha}_n\!\!=\!0|\mathbf{H},K\!\!=\!k,\alpha_n\!\!=\!\!1]]$, and the same average symbol error rate (SER) under receive beamforming strategy $i$ and PSK modulation of size $W$, denoted by 
$\psi^{(i,\!W)}\!(k,\!L)\!\triangleq\! \mathbb{E}_{\mathbf{H}}[\mathrm{Pr}[\hat{s}_n^W\!\!\!\neq\!\! s_n^W|\mathbf{H}, K\!\!=\!k,\hat{\alpha}_n\!\!=\!\alpha_n\!\!=\!\!1]]$, where $n$ represents the index of a typical active user, and $\mathbf{H}\!\triangleq\!({\mathbf{h}_{n}})_{n\in\mathcal{N}}$.
\vspace{-5pt}
\section{Analysis of SSTR}
\vspace{-3pt}
In this section, we derive an approximate expression of the asymptotic $\Phi^{(i,W)}(L,\epsilon)$ at large $M,L,N$ and high SNR. In the regime of $L\leq k$ where AMP does not work, we assume that activity detection and data detection fail, i.e., $p(k,L)=1$ and $\psi^{(i,W)}(k,L)=1$. In the following, we focus on the asymptotic analysis of $p(k,L)$ and $\psi^{(i,W)}(k,L)$ in the regime of $k<L$. First, we use the asymptotic expression of $p(k,L)$ at large $L,N,k$ and high SNR obtained in \textrm{[\citen{liu2018partI}, Theorem 4]} as an approximation for $p(k,L)$ at large $L, N$ and high SNR and $k<L$.
\begin{lemma}\label{le:pk}
	\textrm{[\citen{liu2018partI}, Theorem 4]}
	At large $L$, $N$ and high SNR, for all $k<L$,
	\begin{align}\label{eq:p}
	\setlength{\abovedisplayshortskip}{1pt}
	\setlength{\belowdisplayshortskip}{1pt}
%		\begin{align}{ll}
	&\!\!p(k,\!L)\!\approx\!\frac{\exp{\!(-\!M(b(k,\!L)\!-\!1\!-\!\log{\!(b(k,\!L))}))}}{2\sqrt{2\pi M}}\Big(\!{(1\!-\!b(k,\!L))^{\!-\!1}}\!\nonumber\\
	&\ \ \ \ \ \ \ \ \ +\!{\big(\sqrt{2(b(k,L)\!-\!1\!-\!\log{(b(k,L))})}\big)^{\!-1}} \Big),\!\!
%		\end{align}\!\!
	\end{align}
%	\begin{flalign}\label{eq:p}
%	\setlength{\abovedisplayskip}{3pt}
%	\setlength{\belowdisplayskip}{3pt}
%%	\begin{aligned}
%	&p(k,L)\!\approx\!\frac{\exp{\!(\!-\!M(b(k,L)\!-\!1\!-\!\log{(b(k,L))}))}}{2\sqrt{2\pi M}}\!\big(\!{(1\!-\!b(k,L))^{\!-\!1}}\!\nonumber\\
%	&\ \ \ \ \ \ \ \ +{\big(\sqrt{2(b(k,L)-1-\log{(b(k,L))})}\big)^{-1}} \big),~k<L,\!\!
%%	\end{aligned}
%	\end{flalign}
	where
	$b(k,L)\!\triangleq\!\frac {\sigma^2}{\gamma(L-k)}\log{\Big(\!1\!+\!\frac{\gamma(L-k)}{\sigma^2}\!\Big)}$.
\end{lemma}

Next, we derive an asymptotic approximation of $\psi^{\!(i,\!W)}\!(k,\!L\!)$. 
\begin{lemma}\label{le:ser}
	At large $M,L,N$ and high SNR, for all $k<L$,
	\begin{equation}\label{eq:ser}
	\setlength{\abovedisplayshortskip}{1pt}
	\setlength{\belowdisplayshortskip}{1pt}
	\!\psi^{(i,\!W)}\!(k,\!L)\!\approx\!\!\!\!\begin{aligned}
	&\left\{
	\begin{array}{ll}
	\!\!\!Q\!\left(\!\!\sqrt{2\Gamma^i(k,\!L)}\right),\!\!\!\!&W\!\!=\!2\\
	\!\!\!2Q\!\left(\!\!\sqrt{\Gamma^i(k,\!L)}\right)\!-\!\left(\!Q\!\left(\!\!\sqrt{\Gamma^i(k,\!L)}\right)\!\right)^{\!2}\!\!,\!\!\!\!\!&W\!\!=\!4
	\end{array}\right.\!\!\!\!,\!\!\! 
	\end{aligned}\!\!\!\!\!\!\!\!
	\end{equation}
	where $Q(x)\!=\!\frac 1{\sqrt{2\pi}}\int_{x}^{\infty}\exp\left(\!-\!\frac{t^2}{2}\right)\mathrm{d}t$, and 
	\begin{equation}\label{eq:gamma}
	\setlength{\abovedisplayshortskip}{1pt}
	\setlength{\belowdisplayshortskip}{1pt}
	\Gamma^i(k,L) \!=\! \left\{\!\!\!\!\begin{array}{ll}	\frac{M\gamma^2}{\left(\gamma+ \frac{\sigma^2}{L-k}\right)\left(k\gamma+\sigma^2\right)},&\!\!i=\mathrm{MRC} \vspace{1ex}\\
	\frac{(M-k)(L-k)\gamma^2}{\sigma^2\left(\gamma L+ {\sigma^2}\right)},M>k,&\!\!i=\mathrm{ZF}
	\end{array}\right.\!\!\!.
	\end{equation}
\end{lemma}
\begin{IEEEproof}
	For notation simplicity, let $C_{\!-n}$ denote the event that $\alpha_{n'}\!\!=\!\hat{\alpha}_{n'},n'\!\in\!\mathcal{N},n'\!\neq\! n$. At large $M$, we have:
	\begin{align}
		&\mathrm{Pr}[\hat{s}_n^W\!\!\neq\!\! s_n^W|\mathbf{H}, K\!\!=\!\!k,\hat{\alpha}_n\!\!=\!\!\alpha_n\!\!=\!\!1]\nonumber\\
%			\end{align}
%		\begin{align}
		&\!\!=\mathrm{Pr}[\hat{s}_n^W\!\!\neq\!\! s_n^W|\mathbf{H}, K\!\!=\!\!k,\hat{\alpha}_n\!\!=\!\!\alpha_n\!\!=\!\!1,C_{\!-n}]\mathrm{Pr}[C_{\!-n}|\mathbf{H}, K\!\!=\!\!k,\nonumber\\
		&\ \ \ \hat{\alpha}_n\!\!=\!\!\alpha_n\!\!=\!\!1]\!
		+\!\mathrm{Pr}[\hat{s}_n^W\!\!\neq\!\! s_n^W|\mathbf{H}, K\!\!=\!\!k,
		\hat{\alpha}_n\!\!=\!\!\alpha_n\!\!=\!\!1,\overline{C_{\!-n}}]\nonumber\\		
		&\ \ \times\mathrm{Pr}[\overline{C_{\!-n}}|\mathbf{H},K\!\!=\!\!k,\hat{\alpha}_n\!\!=\!\!\alpha_n\!\!=\!\!1]\nonumber\\
		&\!\!\!\overset{(a)}\approx\mathrm{Pr}[\hat{s}_n^W\!\!\neq\!\! s_n^W|\mathbf{H}, K\!\!=\!\!k,\hat{\alpha}_n\!\!=\!\!\alpha_n\!\!=\!\!1,C_{\!-n}],\nonumber
	\end{align}
	where $(a)$ is due to $\mathrm{Pr}[C_{\!-n}|\mathbf{H}, K\!\!=\!\!k,\hat{\alpha}_n\!\!=\!\!\alpha_n\!\!=\!\!1]\!\!\to\!\! 1$ and $\mathrm{Pr}[\overline{C_{\!-n}}|\mathbf{H}, K\!\!=\!\!k,\hat{\alpha}_n\!\!=\!\!\alpha_n\!\!=\!\!1]\!\!\to\!\! 0$ as $M\!\!\to\!\!\infty$. Accordingly, we assume at large $M$, $\alpha_{n}\!\!=\!\hat{\alpha}_{n},n\!\in\!\mathcal{N}$  \cite{liu2018partI}, \cite{liu2018massive2}. In the following, we analyze $\mathrm{Pr}[\hat{s}_n^W\!\!\neq\!\! s_n^W|\mathbf{H}, K\!\!=\!\!k,\hat{\alpha}_n\!\!=\!\!\alpha_n\!\!=\!\!1,C_{\!-n}]$ as an approximation of 	$\mathrm{Pr}[\hat{s}_n^W\!\!\!\!\neq\!\! s_n^W|\mathbf{H}, K\!\!\!=\!\!k,\hat{\alpha}_n\!\!=\!\!\alpha_n\!\!=\!\!1]$ at large $M$.

    The SINR at a particular channel realization is 
	$
	\widetilde{\Gamma}_n^i(k,L)\!=\!\frac{\rho_{n}^{\mathrm{data}}|\hat{\mathbf{u}}_n^{iH} \hat{\mathbf{h}}_n|^2}{F},
	$
	where $F= \sum_{n'\in\mathcal{N}:{\alpha}_{n'}=1, n'\neq n}\rho_{n'}^{\mathrm{data}}\mathbb{E}[|\hat{\mathbf{u}}_n^{iH}\hat{\mathbf{h}}_{n'}|^2]\!+ \!\!\!\sum_{n'\in\mathcal{N}:{\alpha}_{n'}=1}\rho_{n'}^{\mathrm{data}}\mathbb{E}[|\hat{\mathbf{u}}_n^{iH}\Delta\mathbf{h}_{n'}|^2]\!+\!\mathbb{E}[|\hat{\mathbf{u}}_n^{iH} \mathbf{z}^{\text{data}}|^2]$. 
	At large $M$, in the regime of $k<M$,
	\begin{align}
	&\widetilde{\Gamma}_n^{\mathrm{ZF}}(k,L)\nonumber\\
	&\overset{(b)}=\!\frac{\rho^{\mathrm{data}}_n}{\!\mathbb{E}\!\left[\!\left[\!\left(\!\hat{\mathbf{G}}^H\hat{\mathbf{G}}\!\right)^{\!\!-\!1}\right]_{\!nn}\right]\!\!\left(\sum\limits_{n'\in\mathcal{N}:{\alpha}_{n'}=1}\!\!\!\!\rho_{n'}^{\mathrm{data}}\mathbb{E}\left[\|\Delta\mathbf{h}_{n'}\|^2\right]\!+\!\mathbb{E}[\|\mathbf{z}^{\text{data}}\|^2]\!\right)}\nonumber\\
	&\overset{(c)}\approx\!\frac{(M\!-\!k)(L\!-\!k)\gamma^2}{\sigma^2\left(\gamma L\!+\! {\sigma^2}\right)}\nonumber
	\triangleq\Gamma^{\mathrm{ZF}}(k,L),\nonumber
	\end{align}
	where $(b)$ is due to (\ref{beamforming}), and $(c)$ is due to that $
	\hat{\mathbf{h}}_n \!\sim\! \mathcal{CN}\Big(\mathbf{0},\frac {\rho_n^{\mathrm{pilot}}(L-k)\gamma_n^2}{\rho_n^{\mathrm{pilot}}(L-k)\gamma_n+\sigma^2}\mathbf{I}_M\Big)$
	and 
	$
	\Delta{\mathbf{h}}_n\!\sim\! \mathcal{CN}\Big(\mathbf{0},\frac {\gamma_n\sigma^2}{\rho_n^{\mathrm{pilot}}(L-k)\gamma_n+\sigma^2}\mathbf{I}_M\Big)
	$, as $M\!\to\!\infty$ \cite{liu2018partI},  and $\hat{\mathbf{G}}^H\hat{\mathbf{G}}\!\sim\!\mathcal{W}_k\Big(\!M,\frac {\rho_n^{\mathrm{pilot}}(L-k)\gamma_n^2}{\rho_n^{\mathrm{pilot}}(L-k)\gamma_n+\sigma^2}\mathbf{I}_M\!\Big)$ \cite{ELarsson2013Energy}.
%	\begin{flalign}
%		\widetilde{\Gamma}_n^{\mathrm{ZF}}(k,L)\overset{a}=\frac{\rho^{\mathrm{data}}_n}{\!\mathbb{E}\!\!\left[\!\left[\!\left(\hat{\mathbf{G}}^H\hat{\mathbf{G}}\right)^{\!-1}\right]_{\!nn}\right]\left(\sum\limits_{n'\in\mathcal{N}:{\alpha}_{n'}=1}\!\!\!\!\rho_{n'}^{\mathrm{data}}\mathbb{E}\!\left[\|\Delta\mathbf{h}_{n'}\|^2\right]\!\!+\!\mathbb{E}[\|\mathbf{z}^{\text{data}}\|^2]\right)}
%	\end{flalign}
	At large $M$, 
	\begin{align}
		\widetilde{\Gamma}_n^{\mathrm{MRC}}(k,L)&\overset{(d)}\approx\frac{\rho_{n}^{\mathrm{data}}(\mathbb{E}[|\hat{\mathbf{h}}_n^H \hat{\mathbf{h}}_n|])^2}{D}\nonumber\\
		&\overset{(e)}\approx\frac{M\gamma^2}{\big(\gamma+ \frac{\sigma^2}{L-k}\big)\left(k\gamma+\sigma^2\right)}\nonumber
		\triangleq\Gamma^{\mathrm{MRC}}(k,L),\nonumber
	\end{align}
	where $D=\sum_{n'\in\mathcal{N}:{\alpha}_{n'}=1,n'\neq n}\rho_{n'}^{\mathrm{data}}\mathbb{E}[|\hat{\mathbf{h}}_n^{H}\hat{\mathbf{h}}_{n'}|^2]\!+\! \sum_{n'\in\mathcal{N}:{\alpha}_{n'}=1}\rho_{n'}^{\mathrm{data}}\mathbb{E}[|\hat{\mathbf{h}}_n^{H}\Delta\mathbf{h}_{n'}|^2]\!+\!\mathbb{E}[|\hat{\mathbf{h}}_n^{H} \mathbf{z}^{\text{data}}|^2]\!+\!\rho_{n}^{\mathrm{data}}\mathbb{E}[|\hat{\mathbf{h}}_n^{H}\hat{\mathbf{h}}_{n}|^2]\!-\!\rho_{n}^{\mathrm{data}}(\mathbb{E}[|\hat{\mathbf{h}}_n^{H}\hat{\mathbf{h}}_n|])^2$, $(d)$ is from \cite{marzetta2016fundamentals}, and $(e)$ is due to the distributions of $\hat{\mathbf{h}}_n$ and $\Delta{\mathbf{h}}_n$, as $M\to\infty$ \cite{liu2018partI}.
	Then, by \cite{proakisdigital},
%	$\mathrm{Pr}[\hat{s}_n^W\neq s_n^W|\mathbf{H}, K\!=\!k,\hat{\alpha}_n\!=\!\alpha_n\!=\!1,C_{\!-n}]=$
	$
%	\begin{aligned}
%		\!\!\!\!\!\!\!\!\!\!\!\!\!\!
		\mathrm{Pr}[\hat{s}_n^W\!\!\neq\!\! s_n^W|\mathbf{H}, K\!\!=\!\!k,\hat{\alpha}_n\!\!=\!\!\alpha_n\!\!=\!\!1,C_{\!-n}]\nonumber
		=\nonumber
		\left\{\!\!\!\!
		\begin{array}{ll}
		Q\big(\!\sqrt{2\Gamma^i(k,\!L)}\big),\!\!\!&W=2\\
		2Q\big(\!\sqrt{\Gamma^i(k,\!L)}\big)\!-\!\big(Q\big(\!\sqrt{\Gamma^i(k,\!L)}\big)\!\big)^{\!2}\!\!,\!\!\!\!&W=4\\
		\end{array}	
		\right.
		\!\!\!\!\!\triangleq\!\widetilde{\psi}^{(i,\!W)}\!(k,\!L).
%	\end{aligned}
	$
	As $\psi^{(i,W)}(k,L)\approx\mathbb{E}_{\mathbf{H}}[\mathrm{Pr}[\hat{s}_n^W\neq s_n^W|\mathbf{H}, K\!=\!k,\hat{\alpha}_n\!=\!\alpha_n\!=\!1,C_{\!-n}]]=\widetilde{\psi}^{(i,W)}(k,L)$, we complete the proof.
\end{IEEEproof}

%For MMSE beamforming, assuming $K,M$ go to infinity with $K/M=\mu$, the SINR for user $k$ can be obtained by \cite{liu2018massive2}
%\begin{equation}
%\Gamma^{i}_n\approx\frac{\gamma^2}{\gamma+ \frac{\sigma^2}{L-K}}C,
%\end{equation}
%with $C$ being the unique solution of the fixed point equation:
%$
%\frac MC={ \frac{K\gamma^2}{\gamma+ \frac{\sigma^2}{L-K}+\gamma^2C}+ \frac{K\gamma {\sigma^2}}{\gamma(L-K)+ {\sigma^2}}}.
%$

Based on Lemma \ref{le:pk} and Lemma \ref{le:ser}, we obtain an approximate expression of $\Phi^{(i,W)}(L,\epsilon)$ at large $M$, $N$, $L$ and high SNR.
\begin{theorem}\label{the1}
	At large $M$, $N$, $L$ and high SNR, 
	\begin{equation*} \label{SSTRapp}
	\setlength{\abovedisplayskip}{2pt}
	\setlength{\belowdisplayskip}{1pt}
	\begin{aligned}
	\!\!\!\Phi^{(i,\!W)}(L,\epsilon)\!&\approx\!\!\frac{T\!-\!L}{T} \!\sum_{k=1}^{N}\! kq(k)\!\left(1\!-\!p(k,\!L)\right)\!\left(1\!-\!\psi^{(i,W)}(k,L)\!\right)\\
	&\triangleq\widetilde{\Phi}^{(i,W)}(L,\epsilon),
	\end{aligned}\!\!\!\!
	\end{equation*}
%	\begin{flalign} \label{SSTRapp}
%	\setlength{\abovedisplayskip}{3pt}
%	\setlength{\belowdisplayskip}{3pt}
%%	\begin{aligned}
%	\Phi^{\!(i,W)}(L,\epsilon)\!&\approx\!\nonumber\frac{T\!-\!L}{T} \!\sum_{k=1}^{N}\! kq(k)\!\left(1\!-\!p(k,L)\right)\!\left(\!1\!-\!\psi^{(i,\!W)}\!(k,L)\!\right)\!,\\
%	&\triangleq\widetilde{\Phi}^{(i,W)}(L,\epsilon)
%%	\end{aligned}
%	\end{flalign}
	where $p(k,L)$ is given  by Lemma \ref{le:pk} and $\psi^{(i,W)}(k,L)$ is given by Lemma \ref{le:ser}.
\end{theorem}
\begin{IEEEproof}We have:
	\begin{flalign}
	\setlength{\abovedisplayshortskip}{1pt}
	\setlength{\belowdisplayshortskip}{1pt}
%	\begin{aligned}
	&\Phi^{(i,W)}(L,\epsilon)\!\overset{(a)}=\!\frac{T\!-\!L}{T}N\mathbb{E}_{\mathbf{H}}\left[\mathrm{Pr}[\alpha_n\!=\!1,\hat{\alpha}_n\!=\!1,\hat{s}_n^W\!=\!s_n^W|\mathbf{H}]\right]\nonumber\\
%	\end{flalign}
%	\begin{flalign}
	&=\frac{T\!-\!L}{T}N\sum_{k=1}^{N}\mathbb{E}_{\mathbf{H}}\!\big[\mathrm{Pr}[\alpha_n=1|\mathbf{H}]\mathrm{Pr}[K=k|\mathbf{H},\alpha_n=1]\nonumber\\
	&\times\!\mathrm{Pr}[\hat{\alpha}_n\!\!=\!\!1|\mathbf{H},\!K\!\!\!=\!\!k,\alpha_n\!\!=\!\!1]\mathrm{Pr}[\hat{s}_n^W\!\!\!=\!\!s_n^W|\mathbf{H},\!K\!\!\!=\!\!k,\alpha_n\!\!=\!\!1,\hat{\alpha}_n\!\!=\!\!1]\big]
	\nonumber\\
	&\overset{(b)}=\!\frac{T\!-\!L}{T}N\mathrm{Pr}[\alpha_n\!=\!1]\sum_{k=1}^{N}\mathrm{Pr}[K\!\!=\!k|\alpha_n\!=\!1]\mathbb{E}_{\mathbf{H}}\!\big[\mathrm{Pr}[\hat{\alpha}_n\!=\!1|\mathbf{H},\nonumber\\
	&\ K\!\!=\!k,\alpha_n\!=\!1]\mathrm{Pr}[\hat{s}_n^W\!=\!s_n^W|\mathbf{H},K\!=\!k,\alpha_n\!=\!1,\hat{\alpha}_n\!=\!1]\big]
	\nonumber\\
	&\overset{(c)}\approx\!\frac{T\!-\!L}{T}N\mathrm{Pr}[\alpha_n\!=\!1]\sum_{k=1}^{N}\mathrm{Pr}[K\!=\!k|\alpha_n\!=\!1]\mathbb{E}_{\mathbf{H}}\big[\mathrm{Pr}[\hat{\alpha}_n\!\!=\!\!1|\mathbf{H},\nonumber\\
	&\  K\!\!=\!\!k,\alpha_n\!\!=\!\!1]\big]\mathbb{E}_{\mathbf{H}}\big[\mathrm{Pr}[\hat{s}_n^W\!\!=\!\!s_n^W|\mathbf{H},\!K\!\!=\!\!k,\alpha_n\!=\!\!1,\hat{\alpha}_n\!\!=\!\!1]\big]\!\!\label{eq:proofapp}\\
	&\overset{(d)}=\!\frac{T\!-\!L}{T} \sum_{k=1}^{N} kq(k)\left(1\!-\!p(k,L)\right)\left(1-\psi^{(i,W)}(k,L)\right),\label{prooft}\nonumber
%	\end{aligned}
	\end{flalign} 
%	\begin{flalign}
%	\overset{(e)}=\frac{T-L}{T} \sum_{k=1}^{N} kq(k)\left(1\!-\!p(k,L)\right)\left(1-\psi^{(i,W)}(k,L)\right),\label{prooft}\nonumber
%	\end{flalign}
	where $(a)$ is due to (\ref{eq:def}) and the statistical channel inversion, $(b)$ is due to the independence between $\alpha$, and $\mathbf{H}$, $(c)$ is due that $\mathrm{Pr}[\hat{\alpha}_n\!\!=\!\!1|\mathbf{H},K\!\!=\!\!k,\alpha_n\!\!=\!\!1]$ and $\mathrm{Pr}[\hat{s}_n^W\!\!=\!\!s_n^W|\mathbf{H},K\!\!=\!\!k,\alpha_n\!\!=\!\!\hat{\alpha}_n\!\!=\!\!1]$ become approximately independent at large $M$ \cite{liu2018partI}, and $(d)$ is due to $\mathrm{Pr}[\alpha_n\!\!=\!\!1]\!\!=\!\!p_a\epsilon$, $\mathrm{Pr}[K\!\!=\!\!k|\alpha_n\!\!=\!\!1]\!\!=\!\!q(N\!\!-\!\!1,k\!\!-\!\!1)$, $\mathbb{E}_{\mathbf{H}}[\mathrm{Pr}[\hat{\alpha}_n\!\!=\!\!1|\mathbf{H},K\!\!\!=\!\!k,\alpha_n\!\!=\!\!1]]\!=\!1\!-\!p(k,L)$ and $\mathbb{E}_{\mathbf{H}}[\mathrm{Pr}[\hat{s}_n^W\!\!=\!\!s_n^W|\mathbf{H},\!K\!\!\!=\!\!k,\alpha_n\!\!=\!\!\hat{\alpha}_n\!\!=\!\!1]]\!=\!1\!-\!\psi^{(i,W)}(k,L)$.
\end{IEEEproof}

In Fig. 1, each analytical curve and the corresponding Monte-Carlo points indicate $\frac{T\!-\!L}{T}\left(1\!-\!p(k,L)\right)(1\!-\!\psi^{(i,W)}\!(k,L)\!)$ and $\frac{T\!-\!L}{T}\mathbb{E}_{\mathbf{H}}[\mathrm{Pr}[\hat{\alpha}_n\!=\!\!1|\mathbf{H},K\!\!=\!k,\alpha_n\!=\!\!1]\mathrm{Pr}[\hat{s}_n^W\!\!=\!s_n^W|\mathbf{H},K\!\!=\!k,\alpha_n\!\!=\!\hat{\alpha}_n\!=\!1]]$, respectively. From Fig. 1, we can see that each analytical curve and the corresponding Monte-Carlo points closely match. This demonstrates the accuracy of the approximations in (\ref{eq:proofapp}), Lemma 1 and Lemma 2, and hence demonstrates the accuracy of Theorem 1. In Fig. 2, each analytical curve and the corresponding Monte-Carlo points indicate $\Phi^{(i,W)}(L,\epsilon)$ and $\widetilde{\Phi}^{(i,W)}(L,\epsilon)$, respectively. The fact that each analytical curve and the corresponding Monte-Carlo points closely match further demonstrates the accuracy of Theorem \ref{the1}. The computational complexity for evaluating $\widetilde{\Phi}^{(i,W)}(L,\epsilon)$ is $O(N^3)$.
The closed-form expression $\widetilde{\Phi}^{(i,\!W)}\!(L,\!\epsilon)$ in Theorem 1 can be used for efficiently evaluating and optimizing the SSTR  in practical systems.

From Lemma 1 and Lemma 2, we know that as  $M$ or SNR increases, $p(k,L)$ and $\psi^{(i,W)}(k,L)$ decrease, which results in the increment of $\widetilde{\Phi}^{(i,W)}(L,\epsilon)$.
Other system parameters influence $\widetilde{\Phi}^{(i,W)}(L,\epsilon)$ in very complex manners,  and their impacts have to be obtained using numerical evaluation.
For example, from Fig. 2, we can see that when $N, p_a, \epsilon$ or $L$ is small, $\widetilde{\Phi}^{(i,W)}(L,\epsilon)$ increases with it and when $N, p_a, \epsilon$ or $L$ is large, $\widetilde{\Phi}^{(i,W)}(L,\epsilon)$ decreases with it. The reasons are as follows. 
As $N, p_a$ or $\epsilon$ increases, on average, the number of users sending data (i.e., the number of transmitted data symbols) increases. When $N, p_a$ or $\epsilon$ is small, the accuracy of user activity detection and channel estimation decreases slowly with $N, p_a$ or $\epsilon$, and hence $\widetilde{\Phi}^{(i,W)}(L,\epsilon)$ increases with $N, p_a$ or $\epsilon$. When $N, p_a$ or $\epsilon$ is large, the accuracy of user activity detection and channel estimation decreases fast with $N, p_a$ or $\epsilon$, and hence $\widetilde{\Phi}^{(i,W)}(L,\epsilon)$ decreases with $N, p_a$ or $\epsilon$. 
In addition, a longer pilot length $L$ leads to better user activity detection and channel estimation but fewer transmitted data symbols. 
When $L$ is small, the accuracy of activity detection and channel estimation increases fast with $L$, and hence $\widetilde{\Phi}^{(i,W)}(L,\epsilon)$ increases with $L$. When $L$ is large, the accuracy of activity detection and channel estimation increases slowly with $L$, and hence $\widetilde{\Phi}^{(i,W)}(L,\epsilon)$ decreases with $L$.

\begin{figure}[t]
	\label{verify}
	\centering
	\includegraphics[scale=0.33]{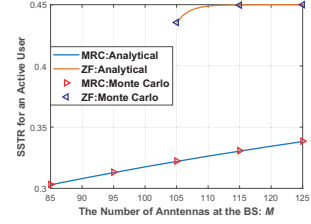}
	\captionsetup{font={footnotesize}}
	\caption{SSTR for an active user at $N=2000, k=100, L=110, T=200,$ SNR $=10 \mathrm{dB}$ and $W=4$.}
	\vspace{-0.3cm}
\end{figure}
\begin{figure}[t]
	\centering
	\captionsetup{font={footnotesize}}
	\subfigure[SSTR versus $N$ at $\epsilon=0.5$, $L=110$, $p_a=0.1$.]{\label{fig_N}
		\includegraphics[scale=0.33]{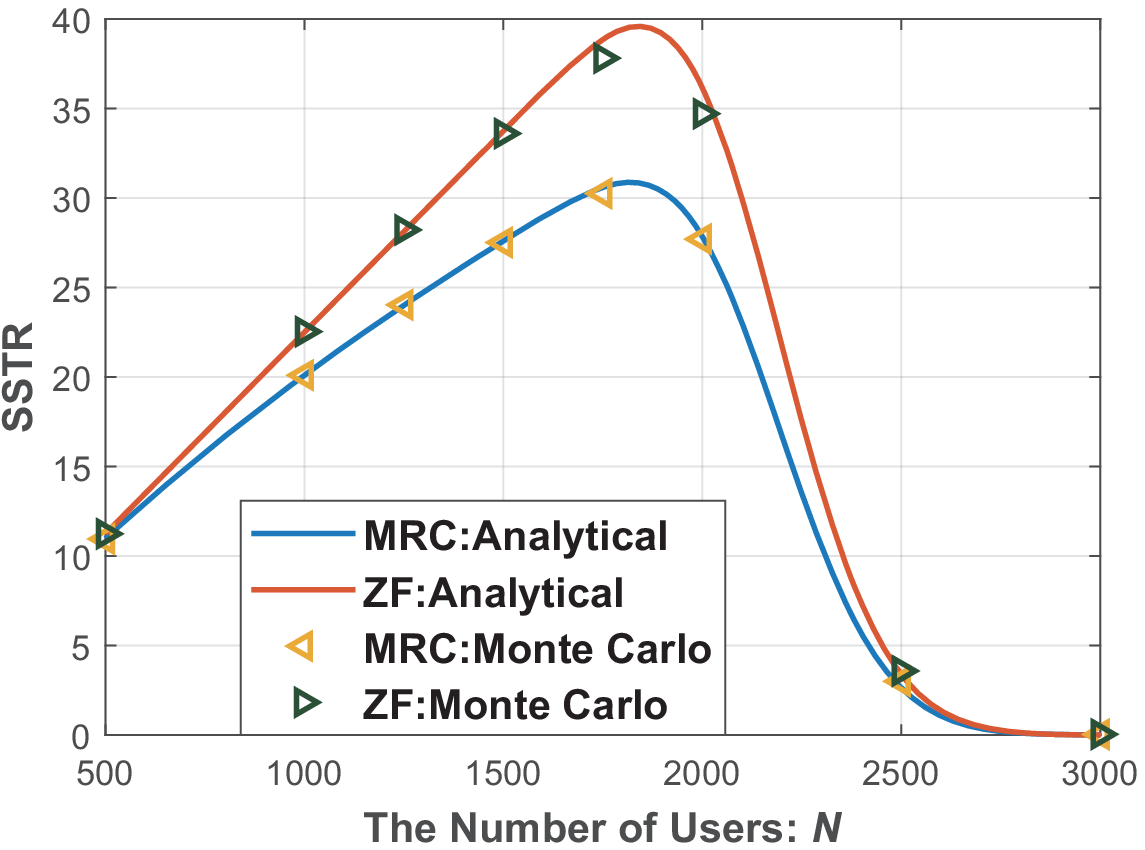}}
	\captionsetup{font={footnotesize}}
	\subfigure[SSTR versus $p_a$ at $\epsilon=0.5$, $L=110$, $N=2000$.]{\label{fig_pa}
		\includegraphics[scale=0.33]{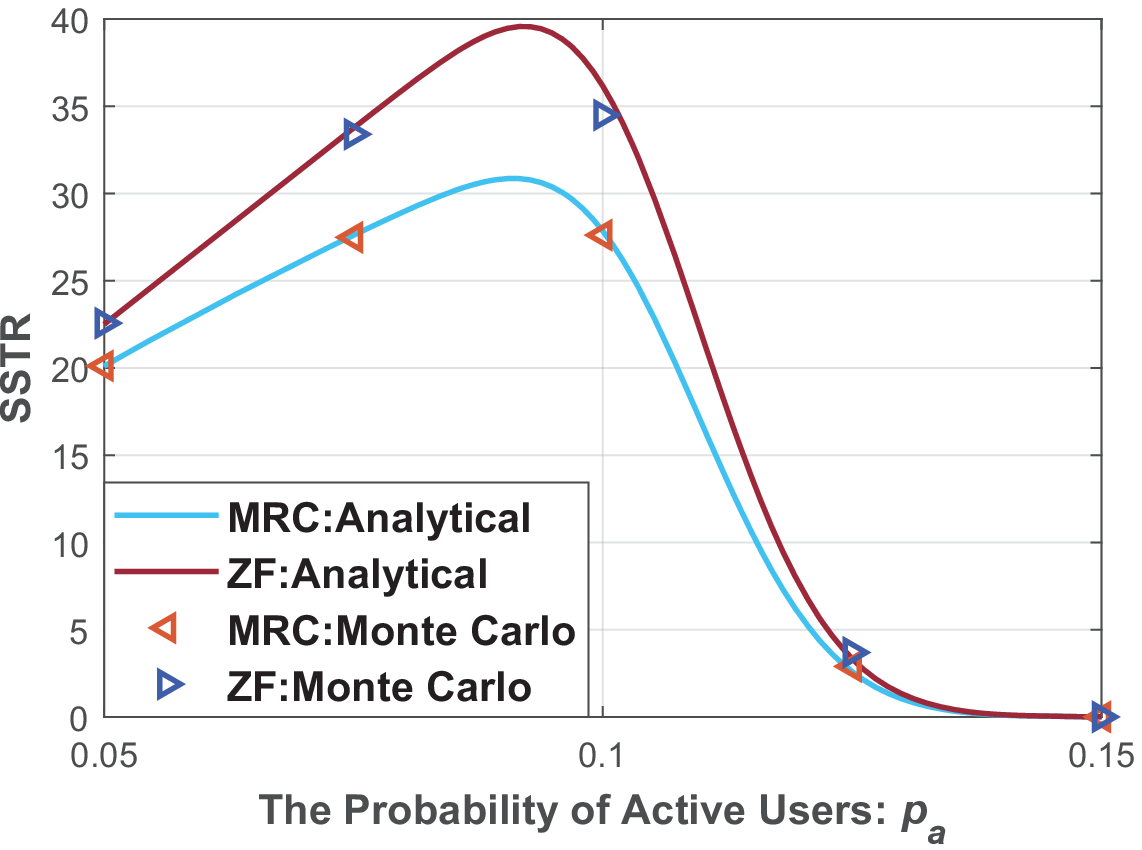}}
	%	\vspace{-0.03in}
	\captionsetup{font={footnotesize}}
	\subfigure[SSTR versus $\epsilon$ at $L=110$, $N=2000, p_a=0.1$.]{\label{sub.1}
		\includegraphics[scale=0.33]{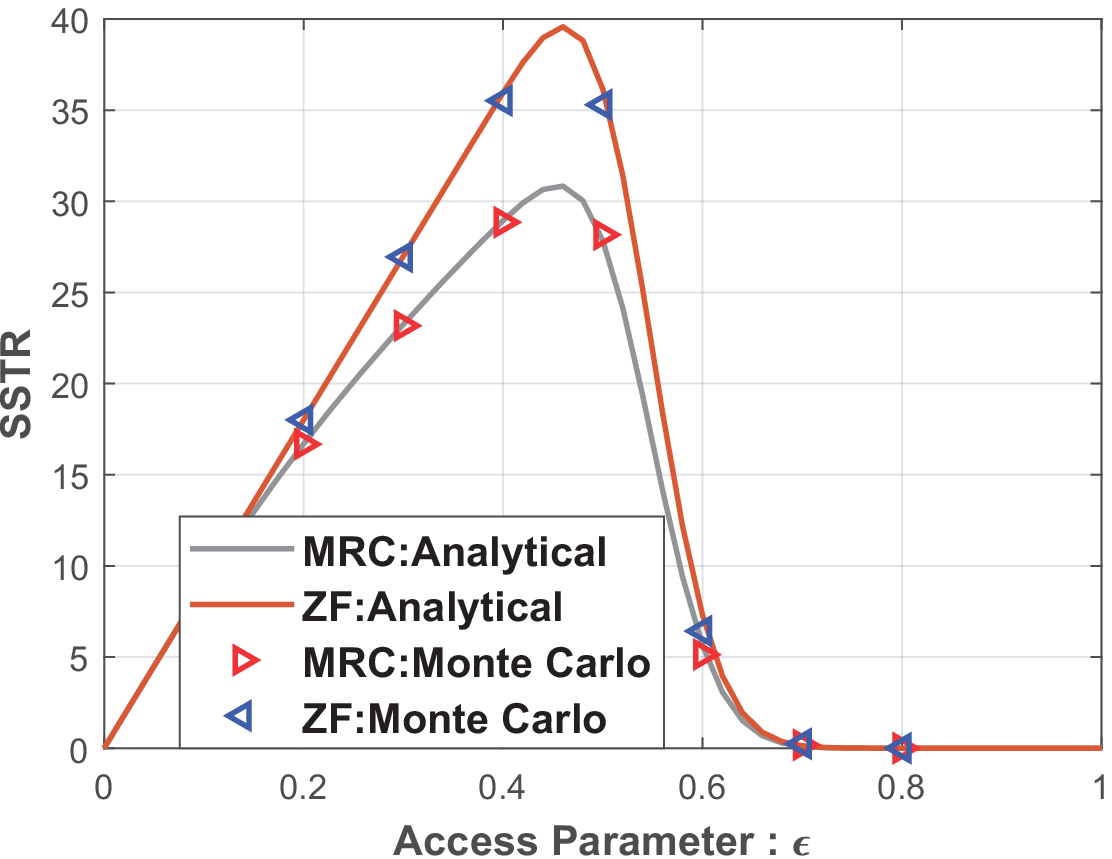}}
	\captionsetup{font={footnotesize}}
	\subfigure[SSTR versus $L$ at $\epsilon=0.5$, $N=2000, p_a=0.1$.]{\label{sub.2}
		\includegraphics[scale=0.33]{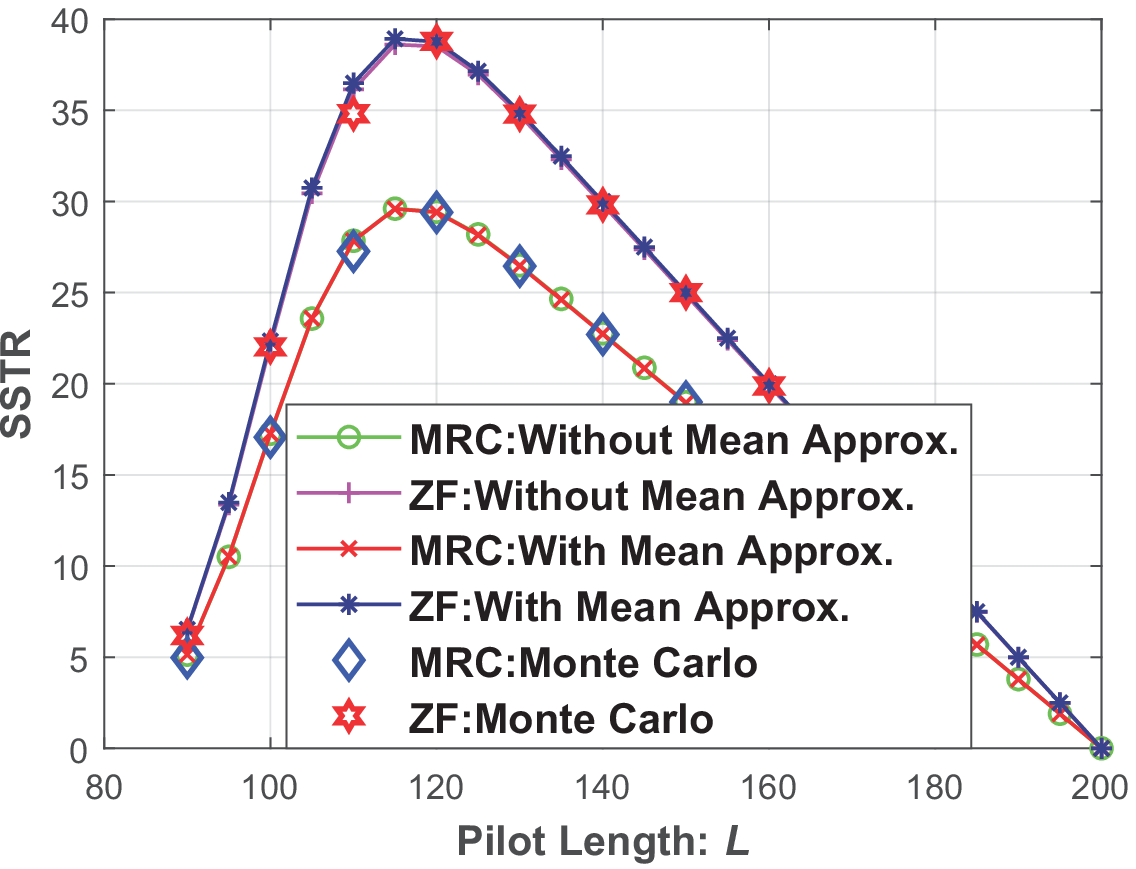}
	}
	\captionsetup{font={footnotesize}}
	\vspace{-0.1cm}
	\caption{SSTR versus $N, p_a, \epsilon$ and $L$ at $M=128, T=200,$  SNR $=10 \mathrm{dB}$ and $W=4$.}
	\vspace{-0.3cm}
\end{figure}

\vspace{-5pt}
\section{Optimization of SSTR}
\vspace{-2pt}
Fig. 2(c) and Fig. 2(d) indicate that it is important to carefully select the system design parameters $\epsilon$ and $L$ so as to improve the SSTR. In this section, we consider the SSTR maximization with respect to $\epsilon$ and $L$.
\vspace{-8pt}
\subsection{Optimization of Access Parameter}
\vspace{-3pt}
In this part, we maximize the SSTR $\widetilde{\Phi}^{(i,W)}(L,\epsilon)$ with respect to $\epsilon$ for given $L$:\footnote{This problem is important for adjusting $\epsilon$ under abnormal conditions (e.g., $p_a$ is far from its typical value).}
\begin{equation}\label{eq:access}
\setlength{\abovedisplayskip}{1pt}
\setlength{\belowdisplayskip}{1pt}
	g(L) \triangleq \max_{0\le\epsilon\le 1}\quad \widetilde{\Phi}^{(i,W)}(L,\epsilon).
	\vspace{-6pt}
\end{equation}
The problem in (\ref{eq:access}) is not in a convex form. By exploiting its structural properties, we have the following result.
\begin{lemma}
	The optimization in (\ref{eq:access}) is equivalent to:
	\begin{equation}\label{eq:maxepsilon}
		\setlength{\abovedisplayskip}{1pt}
		\setlength{\belowdisplayskip}{1pt}
	\begin{aligned}
	g(L)=&\max_{\epsilon,t}\quad \sum_{k=1}^{N} f(k,L)\epsilon^k t^{N-k}\\
	&\ \mathrm{s.t.}\quad\  0\le p_a\epsilon+t\le 1,\ 0\le\epsilon\le 1,
	\end{aligned}
	\end{equation}
	where $f(k,L)\!=\!\frac{T-L}{T}C_N^k{p_a}^k k\left(1\!-\!p(k,L)\right)(1\!-\psi^{(i,W)}(k,L))$.
\end{lemma}
\begin{IEEEproof}
	By Theorem 1, we have:
	\begin{equation*}
	\setlength{\abovedisplayskip}{1pt}
	\setlength{\belowdisplayskip}{1pt}
	\widetilde{\Phi}^{(i,W)}(L,\epsilon)
	=\sum_{k=1}^{N} f(k,L)\epsilon^k (1-p_a\epsilon)^{N-k}.
	\end{equation*}
	By introducing an auxiliary variable $t=1-p_a\epsilon$, the optimization in (\ref{eq:access}) can be equivalently transformed to:
	\begin{equation}
	\setlength{\abovedisplayskip}{1pt}
	\setlength{\belowdisplayskip}{1pt}
	\begin{aligned}\nonumber
	&\max_{\epsilon,t}\quad \sum_{k=1}^{N} f(k,L)\epsilon^k t^{N-k}\\
	&\ \mathrm{s.t.}\quad\   t= 1-p_a\epsilon,\ 0\le\epsilon\le 1.
	\end{aligned}
	\end{equation}
	As $\sum_{k=1}^{N} f(k,L)\epsilon^k t^{N-k}$ is increasing in $t$,  replacing the equality constraint $t=1-p_a\epsilon$ with the inequality constraint $t\le 1-p_a\epsilon$, i.e., $p_a\epsilon+t\le 1$, in the optimization will not change the optimal solution 
	(the inequality constraint is active at the optimal solution). In addition, as $t\!=\!1\!-\!p_a\epsilon$, we can add $t+p_a\epsilon\geq 0$ in the optimization without loss of optimality. Therefore, we complete the proof.
\end{IEEEproof}

The optimization problem in (\ref{eq:maxepsilon}) is a signomial geometric programming (SGP). A stationary point of it can be obtain using complementary geometric programming (CGP) \cite{chiang2005geometric}. We can run CGP multiple times, each with  a random feasible initial point, and choose the stationary point with the largest objective value as a suboptimal solution of the optimization problem in (\ref{eq:maxepsilon}).
We omit the details due to page limitation.
Fig. \ref{sub.1} demonstrates that the optimization with respect to $\epsilon$ for given $L$ is of critical importance for SSTR improvement.

\subsection{Optimization of Pilot Length}
In this part, we maximize the SSTR $\widetilde{\Phi}^{(i,W)}(L,\epsilon)$ with respect to $L$ for given $\epsilon$:{\footnote{This problem is important for the optimization of $L$ without access control.}}
\begin{equation}\label{eq:max}
\max_{L\in\{1,2,\cdots,T-1\}} \widetilde{\Phi}^{(i,W)}(L,\epsilon).
\end{equation}
This  is a discrete optimization problem. Solving it requires computing $\widetilde{\Phi}^{(i,\!W)}\!(L,\!\epsilon)$ (which is a sum of $N$ terms) for all $L\!\in\! \{1,2,\cdots,T\!\!-\!\!1\}$. To reduce computational complexity, we adopt the mean approximation (i.e., approximating the expectation of a function of a random variable by the function of the expectation of the random variable) for $\widetilde{\Phi}^{(i,\!W)}\!(L,\!\epsilon)$:
%	$\widetilde{\Phi}^{(i,W)}(L,\epsilon)=\nonumber$
\begin{flalign}
	&\widetilde{\Phi}^{(i,W)}(L,\epsilon)\nonumber\\
	&\!=\!\left(\sum_{k=1}^{L-1}\!q(k)\!\!\right)\!\!\frac{T\!-\!L}{T} \!\!\sum_{k=1}^{L-1}\!\! \frac{q(k)}{\sum_{k=1}^{L-\!1}\!q(k)}k\!\left(\!1\!\!-\!p(k,\!L)\!\right)\!\!\left(\!1\!\!-\!\psi^{(i,\!W)}\!(k,\!L)\!\!\right)\nonumber\\
	&\!\approx\!\frac{T\!-\!L}{T}\bar{K}_{<L}\!\left(1\!-\!p(\bar{K}_{<L},L)\right)\!\left(1\!-\!\psi^{(i,W)}(\bar{K}_{<\!L},L)\!\right)\!\sum_{k=1}^{L-1}\!q(k)\nonumber\\
	&\!=\!\frac{T\!-\!L}{T}\!\left(\!1\!-\!p(\bar{K}_{\!<\!L},L)\right)\!\left(\!1\!-\!\psi^{(i,W)}(\bar{K}_{\!<\!L},L)\!\right)\!\!\sum_{k=1}^{L-1}\!kq(k),\label{eq:sstrap}\!\!\!
\end{flalign}
where $\bar{K}_{\!<\!L}\triangleq\frac{\sum_{k=1}^{L-1}kq(k)}{\sum_{k=1}^{L-1}q(k)}$.
Given the approximation of $\widetilde{\Phi}^{(i,W)}(L,\epsilon)$ in (\ref{eq:sstrap}), we only need to compute $p(\bar{K}_{\!<\!L},L)$ and $\psi^{(i,W)}(\bar{K}_{\!<\!L},L)$, and find the optimal $L$  for given $\epsilon$ using exhaustive search (i.e., calculate $\widetilde{\Phi}^{(i,W)}(L,\epsilon)$ for all $L\in \{1,2,\cdots,T-1\}$, and select $L$ that achieves the maximum among them). Fig. \ref{sub.2} shows that the error due to mean approximation is negligible. Fig. \ref{sub.2} also demonstrates that the optimization with respect to $L$ for given $\epsilon$ is of great importance for SSTR improvement.
%\vspace{-5pt}
\subsection{Joint Optimization of Pilot Length and Access Parameter}
%\vspace{-3pt}
{In this part, we jointly optimize  $L$ and $\epsilon$ to maximize the SSTR $\widetilde{\Phi}^{(i,W)}(L,\epsilon)$:
\begin{equation}
\label{eq:maxboth}
\max_{0\le\epsilon\le 1,L\in \{1,2,\cdots,T-1\}} \widetilde{\Phi}^{(i,W)}(L,\epsilon),
\end{equation}
which is equivalent to:
\begin{equation}
\max_{L\in \{1,2,\cdots,T-1\}} \max_{0\le\epsilon\le 1,} \widetilde{\Phi}^{(i,W)}(L,\epsilon) = \max_{L\in\{1,2,\cdots,T-1\}} g(L),\nonumber
\end{equation}
where $g(L)$ is given by (\ref{eq:access}).
Thus, we can solve the joint optimization problem in (\ref{eq:maxboth}) based on the optimal solution of the problem in (\ref{eq:access}), and exhaustive search over $L\in \{1,2,\cdots,T-1\}$.
\section{Conclusion}
In this letter, we investigated  grant-free massive access in a massive MIMO system. We considered random access control, and adopted AMP for user activity detection and channel estimation in the pilot transmission phase and PSK modulation in the data transmission phase. We proposed a more reasonable performance metric, i.e., SSTR. We focused on the analysis and optimization of the asymptotic SSTR. Both analysis and optimization results offer important design insights for practical mMTC systems.
\ifCLASSOPTIONcaptionsoff
  \newpage
\fi

\bibliographystyle{IEEEtran}

\begin{thebibliography}{1}
	\providecommand{\url}[1]{#1}
	\csname url@samestyle\endcsname
	\providecommand{\newblock}{\relax}
	\providecommand{\bibinfo}[2]{#2}
	\providecommand{\BIBentrySTDinterwordspacing}{\spaceskip=0pt\relax}
	\providecommand{\BIBentryALTinterwordstretchfactor}{4}
	\providecommand{\BIBentryALTinterwordspacing}{\spaceskip=\fontdimen2\font plus
		\BIBentryALTinterwordstretchfactor\fontdimen3\font minus
		\fontdimen4\font\relax}
	\providecommand{\BIBforeignlanguage}[2]{{%
			\expandafter\ifx\csname l@#1\endcsname\relax
			\typeout{** WARNING: IEEEtran.bst: No hyphenation pattern has been}%
			\typeout{** loaded for the language `#1'. Using the pattern for}%
			\typeout{** the default language instead.}%
			\else
			\language=\csname l@#1\endcsname
			\fi
			#2}}
	\providecommand{\BIBdecl}{\relax}
	\BIBdecl
	
	\bibitem{liu2018partI}
	L.~Liu and W.~Yu, ``Massive connectivity with massive {MIMO}—{P}art {I}:
	Device activity detection and channel estimation,'' \emph{IEEE Trans. Signal
		Process.}, vol.~66, no.~11, pp. 2933--2946, June 2018.
	
	\bibitem{liu2018massive2}
	------, ``Massive connectivity with massive {MIMO}—{P}art {II}: Achievable
	rate characterization,'' \emph{IEEE Trans. Signal Process.}, vol.~66, no.~11,
	pp. 2947--2959, June 2018.
	
	\bibitem{Control}
	Z.~{Sun}, Z.~{Wei} \emph{et~al.}, ``Exploiting transmission control for joint
	user identification and channel estimation in massive connectivity,''
	\emph{IEEE Trans. Commun.}, pp. 1--1, 2019.
	
	\bibitem{NBIoT}
	Y.~E. {Wang}, X.~{Lin}, A.~{Adhikary} \emph{et~al.}, ``A primer on 3gpp
	narrowband internet of things,'' \emph{IEEE Commun. Mag.}, vol.~55, no.~3,
	pp. 117--123, March 2017.
	
	\bibitem{EL2018grant}
	K.~Senel and E.~G. Larsson, ``Grant-{F}ree massive {MTC-E}nabled massive
	{MIMO}: A compressive sensing approach,'' \emph{IEEE Trans. Commun.},
	vol.~66, no.~12, pp. 6164--6175, Dec 2018.
	
	\bibitem{ELarsson2013Energy}
	H.~Q. Ngo, E.~G. Larsson, and T.~L. Marzetta, ``Energy and spectral efficiency
	of very large multiuser {MIMO} systems,'' \emph{IEEE Trans. Commun.},
	vol.~61, no.~4, pp. 1436--1449, April 2013.
	
	\bibitem{marzetta2016fundamentals}
	T.~L. Marzetta and H.~Yang, \emph{Fundamentals of massive MIMO}.\hskip 1em plus
	0.5em minus 0.4em\relax Cambridge University Press, 2016.
	
	\bibitem{proakisdigital}
	G.~Proakis, John \emph{et~al.}, \emph{Digital communications}.\hskip 1em plus
	0.5em minus 0.4em\relax Mc-Graw-Hill, 2001.
	
	\bibitem{chiang2005geometric}
	M.~Chiang \emph{et~al.}, ``Geometric programming for communication systems,''
	\emph{Foundations and Trends{\textregistered} in Communications and
		Information Theory}, vol.~2, no. 1--2, pp. 1--154, 2005.
	
\end{thebibliography}

\end{document}